\documentclass[%
 reprint,
 superscriptaddress,
 amsmath,amssymb,
 aps,
 prl
]{revtex4-1}

\usepackage{graphicx}
\usepackage{dcolumn}
\usepackage{bm}
\usepackage{color}
\usepackage{float}
\usepackage{hyperref}
\usepackage[mathlines]{lineno}

\newcommand{\bfg }{\begin{figure}[htpb]}
\newcommand{\efg }{\end{figure}}
\newcommand{\bt }{\begin{table}[htpb]}
\newcommand{\et }{\end{table}}

\begin{document}

\preprint{APS/123-QED}

\title{Observation of $D_{s}^{\pm}/D^0$ enhancement in Au+Au collisions at $\sqrt{s_{_{\rm NN}}}$ = 200 GeV}

\affiliation{Abilene Christian University, Abilene, Texas   79699}
\affiliation{AGH University of Science and Technology, FPACS, Cracow 30-059, Poland}
\affiliation{Alikhanov Institute for Theoretical and Experimental Physics NRC "Kurchatov Institute", Moscow 117218, Russia}
\affiliation{Argonne National Laboratory, Argonne, Illinois 60439}
\affiliation{American University of Cairo, New Cairo 11835, New Cairo, Egypt}
\affiliation{Brookhaven National Laboratory, Upton, New York 11973}
\affiliation{University of California, Berkeley, California 94720}
\affiliation{University of California, Davis, California 95616}
\affiliation{University of California, Los Angeles, California 90095}
\affiliation{University of California, Riverside, California 92521}
\affiliation{Central China Normal University, Wuhan, Hubei 430079 }
\affiliation{University of Illinois at Chicago, Chicago, Illinois 60607}
\affiliation{Creighton University, Omaha, Nebraska 68178}
\affiliation{Czech Technical University in Prague, FNSPE, Prague 115 19, Czech Republic}
\affiliation{Technische Universit\"at Darmstadt, Darmstadt 64289, Germany}
\affiliation{ELTE E\"otv\"os Lor\'and University, Budapest, Hungary H-1117}
\affiliation{Frankfurt Institute for Advanced Studies FIAS, Frankfurt 60438, Germany}
\affiliation{Fudan University, Shanghai, 200433 }
\affiliation{University of Heidelberg, Heidelberg 69120, Germany }
\affiliation{University of Houston, Houston, Texas 77204}
\affiliation{Huzhou University, Huzhou, Zhejiang  313000}
\affiliation{Indian Institute of Science Education and Research (IISER), Berhampur 760010 , India}
\affiliation{Indian Institute of Science Education and Research (IISER) Tirupati, Tirupati 517507, India}
\affiliation{Indian Institute Technology, Patna, Bihar 801106, India}
\affiliation{Indiana University, Bloomington, Indiana 47408}
\affiliation{Institute of Modern Physics, Chinese Academy of Sciences, Lanzhou, Gansu 730000 }
\affiliation{University of Jammu, Jammu 180001, India}
\affiliation{Joint Institute for Nuclear Research, Dubna 141 980, Russia}
\affiliation{Kent State University, Kent, Ohio 44242}
\affiliation{University of Kentucky, Lexington, Kentucky 40506-0055}
\affiliation{Lawrence Berkeley National Laboratory, Berkeley, California 94720}
\affiliation{Lehigh University, Bethlehem, Pennsylvania 18015}
\affiliation{Max-Planck-Institut f\"ur Physik, Munich 80805, Germany}
\affiliation{Michigan State University, East Lansing, Michigan 48824}
\affiliation{National Research Nuclear University MEPhI, Moscow 115409, Russia}
\affiliation{National Institute of Science Education and Research, HBNI, Jatni 752050, India}
\affiliation{National Cheng Kung University, Tainan 70101 }
\affiliation{Nuclear Physics Institute of the CAS, Rez 250 68, Czech Republic}
\affiliation{Ohio State University, Columbus, Ohio 43210}
\affiliation{Institute of Nuclear Physics PAN, Cracow 31-342, Poland}
\affiliation{Panjab University, Chandigarh 160014, India}
\affiliation{Pennsylvania State University, University Park, Pennsylvania 16802}
\affiliation{NRC "Kurchatov Institute", Institute of High Energy Physics, Protvino 142281, Russia}
\affiliation{Purdue University, West Lafayette, Indiana 47907}
\affiliation{Rice University, Houston, Texas 77251}
\affiliation{Rutgers University, Piscataway, New Jersey 08854}
\affiliation{Universidade de S\~ao Paulo, S\~ao Paulo, Brazil 05314-970}
\affiliation{University of Science and Technology of China, Hefei, Anhui 230026}
\affiliation{Shandong University, Qingdao, Shandong 266237}
\affiliation{Shanghai Institute of Applied Physics, Chinese Academy of Sciences, Shanghai 201800}
\affiliation{Southern Connecticut State University, New Haven, Connecticut 06515}
\affiliation{State University of New York, Stony Brook, New York 11794}
\affiliation{Instituto de Alta Investigaci\'on, Universidad de Tarapac\'a, Arica 1000000, Chile}
\affiliation{Temple University, Philadelphia, Pennsylvania 19122}
\affiliation{Texas A\&M University, College Station, Texas 77843}
\affiliation{University of Texas, Austin, Texas 78712}
\affiliation{Tsinghua University, Beijing 100084}
\affiliation{University of Tsukuba, Tsukuba, Ibaraki 305-8571, Japan}
\affiliation{United States Naval Academy, Annapolis, Maryland 21402}
\affiliation{Valparaiso University, Valparaiso, Indiana 46383}
\affiliation{Variable Energy Cyclotron Centre, Kolkata 700064, India}
\affiliation{Warsaw University of Technology, Warsaw 00-661, Poland}
\affiliation{Wayne State University, Detroit, Michigan 48201}
\affiliation{Yale University, New Haven, Connecticut 06520}

\author{J.~Adam}\affiliation{Brookhaven National Laboratory, Upton, New York 11973}
\author{L.~Adamczyk}\affiliation{AGH University of Science and Technology, FPACS, Cracow 30-059, Poland}
\author{J.~R.~Adams}\affiliation{Ohio State University, Columbus, Ohio 43210}
\author{J.~K.~Adkins}\affiliation{University of Kentucky, Lexington, Kentucky 40506-0055}
\author{G.~Agakishiev}\affiliation{Joint Institute for Nuclear Research, Dubna 141 980, Russia}
\author{M.~M.~Aggarwal}\affiliation{Panjab University, Chandigarh 160014, India}
\author{Z.~Ahammed}\affiliation{Variable Energy Cyclotron Centre, Kolkata 700064, India}
\author{I.~Alekseev}\affiliation{Alikhanov Institute for Theoretical and Experimental Physics NRC "Kurchatov Institute", Moscow 117218, Russia}\affiliation{National Research Nuclear University MEPhI, Moscow 115409, Russia}
\author{D.~M.~Anderson}\affiliation{Texas A\&M University, College Station, Texas 77843}
\author{A.~Aparin}\affiliation{Joint Institute for Nuclear Research, Dubna 141 980, Russia}
% ...

\collaboration{STAR Collaboration}\noaffiliation

\date{\today}

\begin{abstract}
We report on the first measurement of charm-strange meson $D_s^{\pm}$ production at midrapidity 
in Au+Au collisions at $\sqrt{s_{_{\rm NN}}}$ = 200 GeV from the STAR experiment.
The yield ratio between strange ($D_{s}^{\pm}$) and non-strange ($D^{0}$) open-charm mesons is presented and compared to model calculations.
A significant enhancement, relative to a PYTHIA simulation of $p$+$p$ collisions, is observed in the $D_{s}^{\pm}/D^0$ yield ratio in Au+Au collisions over a large range of collision centralities. Model calculations incorporating abundant strange-quark production in the quark-gluon plasma (QGP) and coalescence hadronization qualitatively reproduce the data.
The transverse-momentum integrated yield ratio of $D_{s}^{\pm}/D^0$ at midrapidity is consistent with a prediction from a statistical hadronization model with the parameters constrained by the yields of light and strange hadrons measured at the same collision energy.
These results suggest that the coalescence of charm quarks with strange quarks in the QGP plays an important role in $D_{s}^{\pm}$ meson production in heavy-ion collisions.
\end{abstract}

\maketitle


At extremely high temperatures and energy densities, a new state of matter in which quarks and gluons are the degrees of freedom, the quark-gluon plasma (QGP), is formed~\cite{QGPth,QGPexp}. 
Since the masses of charm and bottom quarks are larger than the typical temperature ($\sim$300 MeV)~\cite{QGPTem1,QGPTem2} of the QGP formed at the Relativistic Heavy Ion Collider (RHIC), heavy quarks are predominantly produced via initial hard scatterings, and their production cross sections can be evaluated by perturbative quantum chromodynamics (pQCD)~\cite{PQCD1,PQCD2}.

Charm quarks are produced on time scales shorter than the QGP formation, and they subsequently experience the whole evolution of the QGP matter. 
With thermal relaxation time ($\sim$5-10 fm/$c$)~\cite{time1} comparable to the QGP lifetime~\cite{time2,time3}, they carry information about the transport properties of the medium.
During the cooling down of the medium, the charm quarks can hadronize into different open-charm hadrons, $e.g.$, $D^0$, $D^{\pm}$, $D_{s}^{\pm}$ and $\Lambda_{c}^{\pm}$.  How these open-charm hadrons are formed is of particular interest.
In $p$+$p$/$e$+$e$/$p$+$e$ collisions, charm-hadron production at high transverse momentum ($p_{T}$) is well described by the PYTHIA event generator~\cite{PYTHIA} in which the transition of charm quarks into hadrons is described by fragmentation models, such as the Lund string model~\cite{frag1,frag2}.  
In the QGP medium, one expects a different hadronization mechanism through the recombination of charm quarks and light/strange quarks (namely coalescence hadronization)~\cite{coal1,coal2,coal3,coal4, coal5} to dominate at low $p_{T}$ ($<$~5 GeV/c) and fragmentation hadronization to dominate at higher $p_{T}$. 
Support for the coalescence hadronization picture in the charm sector has been observed in a recent STAR measurement of $\Lambda_{c}^{\pm}$ baryon production in Au+Au collisions at $\sqrt{s_{_{\rm NN}}}$ = 200 GeV~\cite{Lambda_c}.

Strange quarks are abundantly produced in the QGP, and (because their mass is comparable to the medium temperature) they are in chemical equilibrium with the fireball~\cite{SEth,SEexp1,SEexp2,SEexp3}.
An increased $D_{s}^{\pm}$ production in heavy-ion collisions relative to $p+p$ collisions has been predicted in case of hadronization via quark recombination due to the enhanced strange-quark abundance in the QGP~\cite{coal5,TAMU_old}.
The $D^{0}$ $p_{T}$ spectra have been measured previously by STAR~\cite{HFTD0} exploiting the high precision of the Heavy Flavor Tracker (HFT)~\cite{MAPS}. 
These results provide a good reference for the study of $D_{s}^{\pm}$ enhancement through comparison of yields of $D_{s}^{\pm}$ and $D^{0}$ mesons as a function of $p_{T}$ for different collision centralities. 
Comparing the $D_{s}^{\pm}$/$D^{0}$ yield ratio in heavy-ion collisions with that in $p+p$ collisions 
helps us understand the QGP effects on charm-quark hadronization.
Various $D_{s}^{\pm}$ measurements have been carried out by the LHC experiments~\cite{LHCexp1,LHCexp2,LHCexp3,LHCexp4,LHCexp5,LHCexp6,LHCexp7}.
Those measurements suggest a possible enhancement of the $D_{s}^{\pm}$/$D^{0}$ yield ratio in Pb+Pb collisions compared to $p+p$ collisions, but the uncertainties are large and prevent a firm conclusion.


\begin{figure}[htbp]
\center{
\includegraphics[width=0.9\columnwidth]{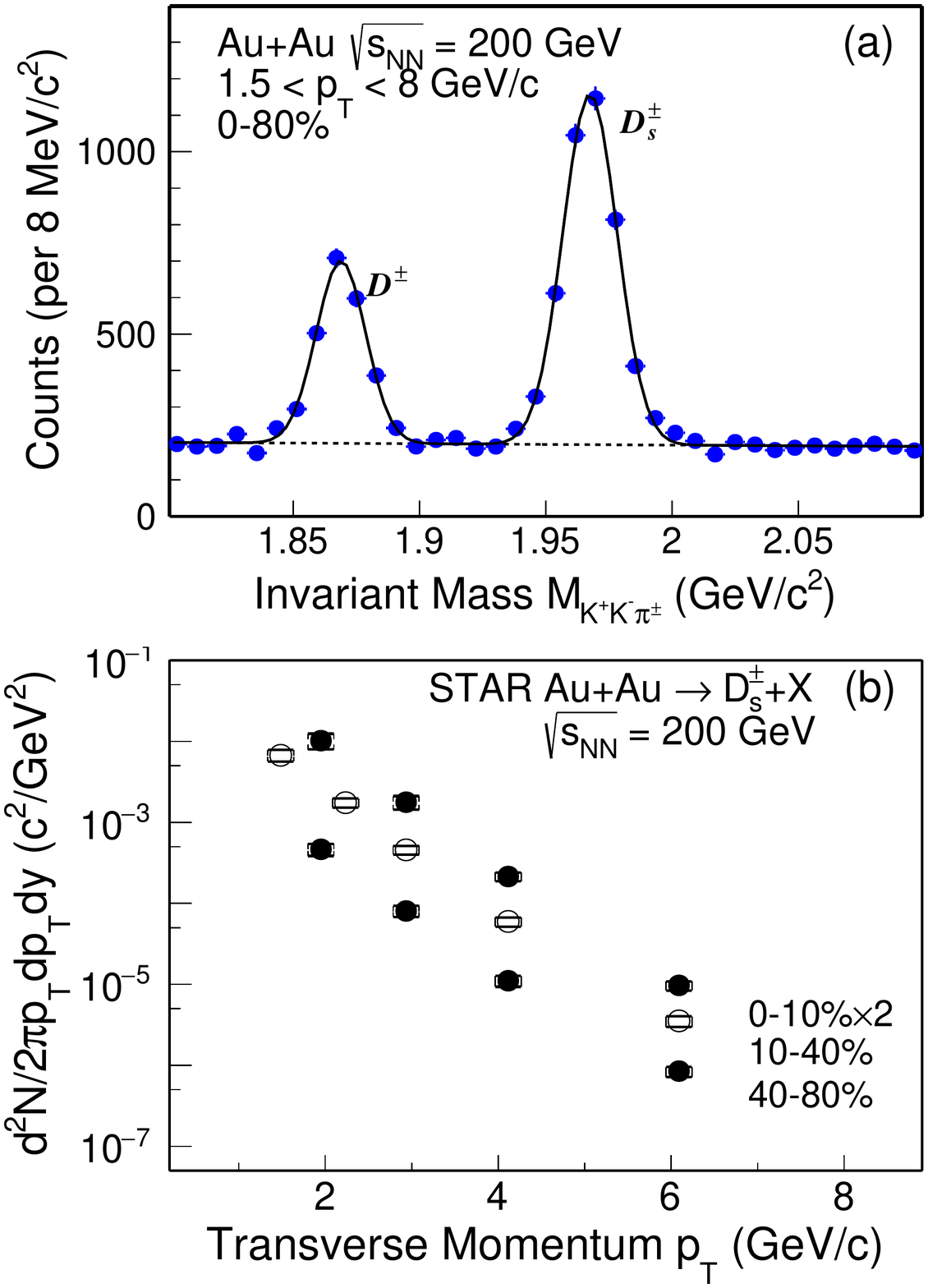}
}
\caption{(a) Invariant mass distribution $M_{K^+K^-\pi^{\pm}}$ of $D_{s}^{\pm}$ candidates in 0--80\% Au+Au collisions at $\sqrt{s_{_{\rm NN}}}$ = 200\,GeV. The solid line depicts a fit with two Gaussian functions representing $D_{s}^{\pm}$ and $D^{\pm}$ signal plus a linear function for background. (b) $D_s^{\pm}$ invariant yield as a function of $p_{T}$ in various centrality intervals of Au+Au collisions at $\sqrt{s_{_{\rm NN}}}$ = 200\,GeV. Vertical bars and brackets on data points represent statistical and systematic uncertainties, respectively.}
\label{fig:DsInvmass_Sepctrum}
\end{figure}

In this letter, we report on the first measurement of $D_{s}^{\pm}$ production over the transverse momentum range of 1\,$<$\,$p_{T}$\,$<$\,8\,GeV/$c$, in Au+Au collisions at a center-of-mass energy of $\sqrt{s_{_{\rm NN}}}$ = 200\ GeV.
The measurement was performed via invariant-mass reconstruction of the hadronic decay channel, $D_{s}^{+}\rightarrow \phi+\pi^+ \rightarrow K^{+} + K^{-}+\pi^+$ (branching ratio (2.24 $\pm$ 0.08)\%) and its charge conjugate~\cite{PDG}.
Approximately 2 billion minimum-bias triggered events recorded by the STAR experiment at RHIC in years 2014 and 2016 are used for this analysis.
The STAR subsystems used in this analysis are immersed in a 0.5 Tesla uniform magnetic field along the beam axis.
The HFT detector~\cite{MAPS} was used to better distinguish open heavy-flavor particles via their decay topologies. The HFT is comprised of three subsystems: the innermost two layers of the Pixel detectors (PXL)~\cite{MAPS}, the Intermediate Silicon Tracker (IST) and the outermost layer of the Silicon Strip Detector (SSD)~\cite{SSD}.
The TPC~\cite{TPC} and the Time-Of-Flight (TOF)~\cite{TOF}, a multi-gap resistive plate chamber, are used to identify charged particles. 
The events used for charm-hadron reconstruction were required to have a primary vertex located along the beamline within 6 cm from the center of the HFT to ensure good HFT acceptance.
A maximum difference of 3 cm between the primary vertex positions reconstructed with the TPC and the Vertex Position Detectors (VPD)~\cite{VPD} was also required in the event selection to reject out-of-time pileup events. 
Collision centrality is determined using the measured charged-particle multiplicity within pseudorapidity $|\eta|$\,$<$\,0.5 and comparing it to a Monte Carlo Glauber simulation~\cite{Cen}.
The tracks used in the $D_{s}^{\pm}$ meson reconstruction are those with at least 20 hits recorded by the TPC, one hit in each PXL layer, and at least one hit in either the IST or SSD.
Those tracks must also be within $|\eta|$\,$<$\,1 and above a minimum $p_{T}$ of 0.5 (0.6)\,GeV/$c$ for kaons (pions). The distance of closest approach (DCA) of the tracks to the primary vertex is required to be larger than 60 $\mu$m in order to reduce the combinatorial background.
The kaons (pions) are identified by selecting tracks within 2 (3) standard deviations of the measured ionization energy loss in the TPC ($dE/dx$) relative to the theoretical value~\cite{Bischel}.
If TOF information is available, $1/\beta$ ( where $\beta$ is the flight velocity of the particle) is additionally required to be less than 3 standard deviations relative to the theoretical value.

The lifetime of $\phi$ mesons is $\sim$50 fm/$c$. Experimentally, the $D_{s}^{\pm}$ mesons are regarded as decaying into $K^{+}K^{-}\pi^{\pm}$ at a single secondary vertex.
The invariant mass of $K^{+}K^{-}$ pairs is required to be within 1.011-1.027 GeV/$c^{2}$ for selecting $\phi$ candidates.    
To improve the significance of the reconstructed $D_{s}^{\pm}$, a machine learning algorithm, the Boosted Decision Tree (BDT) from the Toolkit for MultiVariate Analysis (TMVA)~\cite{TMVA} was employed.
The BDT classifier was obtained by training the signal sample from a data-driven simulation (described elsewhere~\cite{HFTD0}) and a background sample from wrong-sign combinations of $KK\pi$ triplets.
The BDT classifiers obtained using background samples from side bands and wrong-sign combinations were found to give consistent results.
The variables characterizing the $D_{s}^{\pm}$ decay topology such as the DCA of the decay-particle tracks to the primary vertex, the DCA between decay daughters, and the decay length were used as input variables to the BDT classifier. 
The selection on the BDT response was optimized to have the best signal significance based on the number of signal and background counts expected in data.
Figure 1 (a) shows the invariant mass distribution of $K^+K^-\pi^{\pm}$ triplets in the 0--80\% centrality interval. 
The solid line depicts a fit with two Gaussian functions representing the $D_{s}^{\pm}$ and $D^{\pm}$ signals plus a linear function for the background.
The raw-signal yields are obtained by counting the $D_{s}^{\pm}$ candidates in the invariant-mass distribution within 3 standard deviations from the mean of the Gaussian fit, and subtracting the combinatorial background calculated by integrating the linear fit function within the same range.
The mean of the Gaussian function is consistent with the Particle Data Group (PDG)~\cite{PDG} value of the $D_{s}^{\pm}$ mass, and the width is consistent with a Monte Carlo (MC) simulation that includes the momentum resolution.
The raw-signal yields contain the promptly produced $D_{s}^{\pm}$ and the non-prompt $D_{s}^{\pm}$ from $B$-meson decays that satisfy the topological selections. 

The efficiency of $D_{s}^{\pm}$ reconstruction is evaluated via the data-driven simulation validated in the $D^{\rm 0}$ spectra measurement with the HFT~\cite{HFTD0}.
The $D_{s}^{\pm}$ mesons are generated via MC simulation with uniform rapidity, and $p_{T}$ distributions weighted according to the $D^{0}$ yields, and the decay kinematics from the PYTHIA package (version 8.2, Monash tune)~\cite{PYTHIA,Monash}.
The efficiency includes: the acceptance ($|\eta|$ $<$ 1), track $p_{T}$ and quality selection criteria, TPC-to-HFT matching, particle identification (PID), and topological selections.
The impact of the finite primary-vertex resolution on the reconstruction efficiency was estimated with a similar procedure as in Ref.~\cite{HFTD0} based on HIJING~\cite{HIJING} + GEANT3~\cite{GEANT} simulations.
The reconstruction efficiency ($10^{-4}$-$10^{-2}$) for the $D_{s}^{\pm}$ is lower compared to the $D^{0}$ one, and it decreases from peripheral to central collisions and increases with increasing $p_{T}$.
The lower efficiency at low $p_{T}$ and for central collisions is because of lower tracking/vertexing efficiencies for TPC/HFT and more stringent selections applied in the analysis.
The $D_{s}^{\pm}$ invariant yield, ((1/2$\pi p_{T}$)$d^{2}N$/$dp_{T}dy$), is calculated for each centrality and $p_{T}$ interval as the raw signal per event averaged between particles and antiparticles ($N_{(D_{s}^{+} + D_{s}^{-})}$/2/$N_{\rm evt}$), scaled by the inverse of the reconstruction efficiency and the decay branching ratio from the PDG~\cite{PDG}.

The systematic uncertainties have contributions due to the raw-yield extraction, the efficiency calculation, and the feed-down from bottom-hadron decays.
The systematic uncertainty on the raw yield was calculated to be 2-10\%, depending on $p_{T}$ and centrality, by changing the fitting ranges and function types for the background estimate.
The systematic uncertainties due to the track reconstruction efficiency in the TPC and the PID were evaluated by varying the selection criteria, and they were estimated to be $\sim$9\% and $\sim$3\%, respectively, for $KK\pi$ triplets.
The uncertainty on TPC-to-HFT matching efficiency was estimated to be $\sim$3\%~\cite{HFTD0} for $KK\pi$ triplets.
The uncertainty on the topological selection efficiency is determined to be 2-20\%, which is estimated by varying the BDT selection criterion to adjust the efficiency by $\sim$$\pm$50\% relative to the optimized one~\cite{HFTD0}.
A systematic uncertainty of 1-20\% originates from the choice of the generated $D_{s}^{\pm}$ $p_{T}$ spectrum used to determine the efficiency.
This was estimated by comparing the difference in reconstruction efficiency evaluated using the $D_{s}^{\pm}$ and $D^{0}$ spectra. The uncertainty is larger at higher $p_{T}$ and for more peripheral collisions.
 
The non-prompt $D_{s}^{\pm}$ yield (feed-down from $B$-meson decays) was estimated by taking the $B$-hadron spectra from a pQCD FONLL calculation~\cite{FONLL1,FONLL2}, scaling them to an expectation in Au+Au collisions (taking into account the collision geometry and the suppression in the medium from a model calculation~\cite{BRAA}), and then processing them through the data-driven simulation with the full analysis procedure.
A possible $B^{0}_{s}$ enhancement~\cite{Bs} was not considered in this estimate.
The feed-down contribution is evaluated to be 2\% at $p_{T}\sim$2.5\,GeV/$c$ and increases to 10\% at $p_{T}\sim$6\,GeV/$c$.
The feed-down contribution is not subtracted, and it is regarded as an asymmetric systematic uncertainty in the yields and ratios.
In the $D_{s}/D^{0}$ yield ratio, the feed-down contribution partially cancels, leaving 2-6\% contribution at 2.5 $<$ $p_{T}$ $<$ 8\,GeV/$c$, and it is less than 1\% in the lower $p_{T}$ region.
 
The final systematic uncertainty is the sum in quadrature of the contributions from the different sources. 
Finally, the uncertainty from the decay branching ratio is considered as a global normalization uncertainty ($\sim$3.5\%)~\cite{PDG}.
\begin{figure}[htbp]
\center{
\includegraphics[width=0.9\columnwidth]{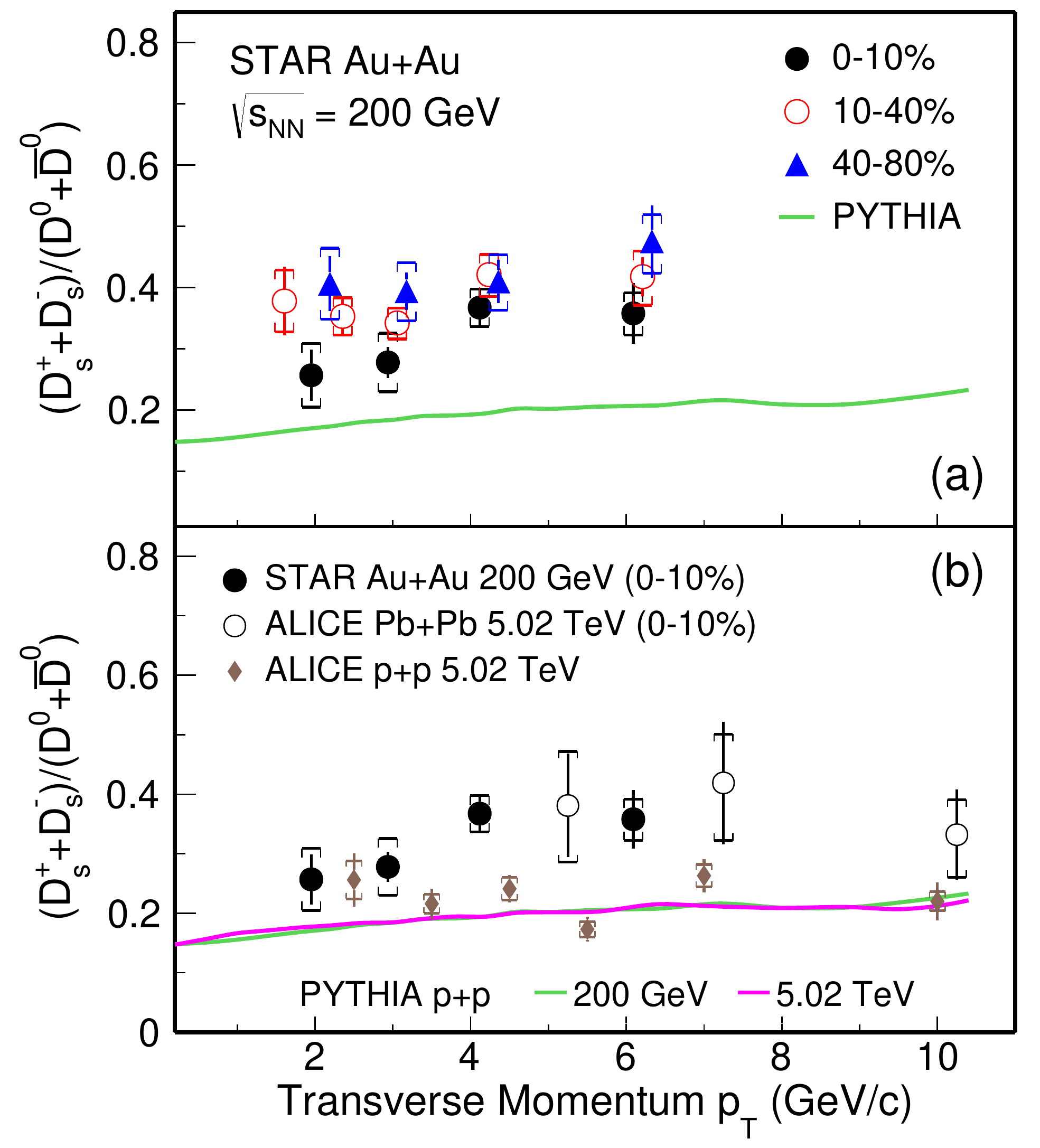}
}
\caption{(a) $D_{s}/D^{0}$ yield ratio as a function of $p_{T}$ in various centrality intervals of Au+Au collisions at $\sqrt{s_{_{\rm NN}}}$ = 200\,GeV, compared to a PYTHIA (version 8.2, Monash tune) simulation for $p$+$p$ collisions at the same energy. (b) STAR measurement of $D_{s}/D^{0}$ yield ratio (black solid points) as a function of $p_{T}$ in 0--10\% central Au+Au collisions at $\sqrt{s_{_{\rm NN}}}$ = 200\,GeV, compared to ALICE measurements in Pb+Pb collisions at $\sqrt{s_{_{\rm NN}}}$ = 5.02\,TeV (open circles) and in $p$+$p$ collisions at $\sqrt{s}$ = 5.02\,TeV (solid diamonds), as well as to PYTHIA simulations for $p$+$p$ collisions at 200\,GeV and 5.02\,TeV (green and purple curves). Vertical bars and brackets on data points represent statistical and systematic uncertainties, respectively.}
\label{fig:DsD0data}
\end{figure}

The invariant yields of $D_{s}^{\pm}$ as a function of $p_{T}$ in different centrality intervals are shown in Fig. 1 (b).   
The statistical and systematic uncertainties on the data points are denoted by vertical bars (smaller than the marker size when not visible) and brackets, respectively.
The ratios of the invariant yield of $D_s^{\pm}$ over that of $D^{0}$ as a function of $p_{T}$ in Au+Au collisions at $\sqrt{s_{_{\rm NN}}}$ = 200\,GeV are shown in Fig. 2.
The correlated systematic uncertainties from the tracking efficiency correction going into both $D_s^{\pm}$ and $D^{0}$ partially cancel in the ratio.
Figure 2 (a) shows the $D_{s}/D^{0}$ yield ratio as a function of $p_{T}$ for different collision centralities compared to that from a PYTHIA (version 8.2, Monash tune) simulation of $p$+$p$ collisions at the same energy.
It is observed that the $D_{s}/D^{0}$ ratio in Au+Au collisions shows a large enhancement (about 1.2-2 times) relative to the PYTHIA simulation of $p$+$p$ collisions, and there is no centrality/$p_{T}$ dependence within the uncertainties.
For the 10--40\% centrality interval,  the significances of the enhancement are 3.8, 5.6, 5.6, 6.0 and 4.6 standard deviations from the first to the last $p_{T}$ bin, respectively.
This indicates that the hadronization of charm quarks is different in heavy-ion collisions compared to $p$+$p$ collisions. 

Figure 2 (b) compares the present STAR results with the $D_{s}/D^{0}$ yield ratio from the ALICE collaboration in Pb+Pb collisions at $\sqrt{s_{_{\rm NN}}}$ = 5.02\,TeV (open circles) in the 0--10\% centrality interval~\cite{LHCexp1} and $p$+$p$ collisions at $\sqrt{s}$ = 5.02\,TeV~\cite{ALICEppDs} (solid diamonds).
It shows that the ratio measured in $p$+$p$ collisions at the LHC is well described by PYTHIA simulations that predict the same $D_{s}/D^{0}$ ratio at the two collision energies. 
STAR measurements in Au+Au collisions are compatible within uncertainties with the ALICE results~\cite{LHCexp1} in Pb+Pb collisions at higher $\sqrt{s_{_{\rm NN}}}$ = 5.02\,TeV in the overlapping $p_{T}$ region for the 0--10\% centrality interval.

\begin{figure}[htbp]
\center{
\includegraphics[width=0.9\columnwidth]{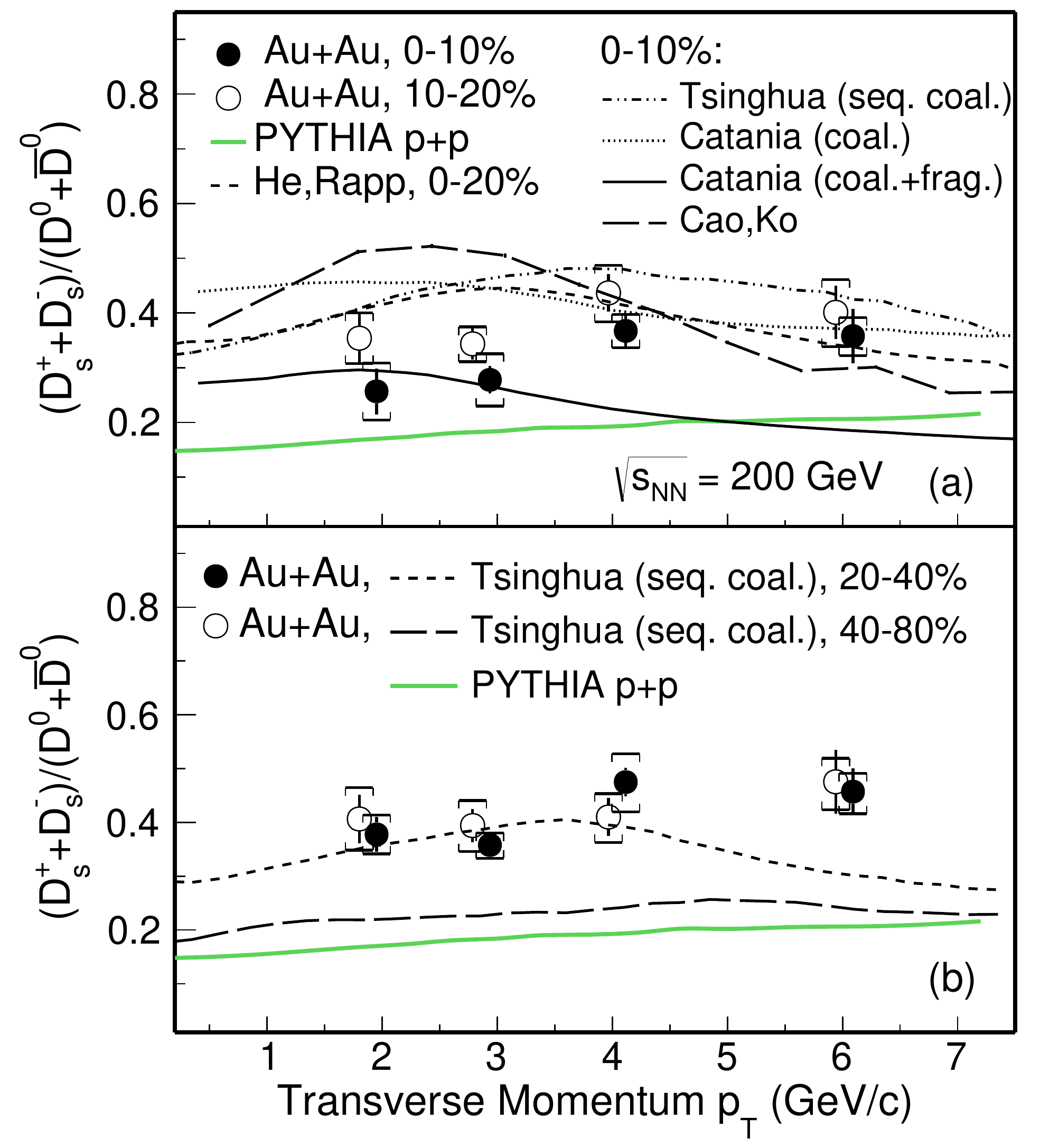}
}
\caption{(a) $D_{s}/D^{0}$ yield ratio as a function of $p_{T}$ compared to various model calculations from He/Rapp (0--20\%), Tsinghua, Catania and Cao/Ko in 0--10\% centrality interval of Au+Au collisions, and PYTHIA prediction in $p$+$p$ collisions at $\sqrt{s_{_{\rm NN}}}$ = 200\,GeV. (b) $D_{s}/D^{0}$ yield ratio as a function of $p_{T}$ compared to model calculations from Tsinghua in 20--40\% (solid circles) and 40--80\% (open circles) centrality intervals of Au+Au collisions at $\sqrt{s_{_{\rm NN}}}$ = 200\,GeV. Vertical bars and brackets on data points represent statistical and systematic uncertainties, respectively.}
\label{fig:DsD0model}
\end{figure}

Figure 3 shows the $D_{s}/D^{0}$ yield ratio as a function of $p_{T}$, for different collision centralities, compared to models that include a contribution to hadronization via coalescence.
These models assume that $D_s^{\pm}$ mesons can be formed by the recombination of charm quarks with strange quarks in the QGP.
Different from the others, the Tsinghua model~\cite{Tsinghua} implements a sequential coalescence ($D_{s}^{\pm}$ mesons hadronize earlier than $D^{0}$) which results in a further enhancement of the $D_{s}/D^{0}$ ratio.
The calculations from `Tsinghua (seq. coal.)' and `Catania (coal.)'~\cite{Catania} include only coalescence hadronization.
The calculations from `Catania (coal.+frag.)', `He/Rapp'~\cite{TAMU} and `Cao/Ko'~\cite{CaoKo} include both coalescence and fragmentation in their modeling of the charm-quark hadronization.
For the most central collisions (0--10/20\%), the predictions from He/Rapp, Cao/Ko, Catania (coal.) and Tsinghua (seq. coal.) generally describe the measured $D_{s}/D^{0}$ ratio. 
The Catania (coal+frag) model describes well the measured $D_{s}/D^{0}$ ratio for $p_{T}$ $<$ 3 GeV/$c$, while at higher $p_{T}$ it underestimates the data and for $p_{T}$ $>$ 4 GeV/$c$ it does not show any enhancement of the ratio relative to PYTHIA simulations.
The predictions of the Tsinghua model for the 20--40\% and 40--80\% centrality intervals are compared to the measured $D_{s}/D^{0}$ ratio in the bottom panel of Fig. 3. 
In the Tsinghua calculations, the $D_{s}/D^{0}$ ratio is driven by the degree of charm-quark thermalization, and therefore it decreases from central to peripheral collisions, reaching a value close to the results of the PYTHIA simulations of $p+p$ collisions in the 40--80\% centrality class.
The model describes the data well for $p_{T}$ $<$ 4 GeV/$c$ in the 20--40\% centrality class, while it significantly underestimates the data in the 40-80\% interval.
Overall, these comparisons indicate that coalescence hadronization plays an important role in charm-quark hadronization in the surrounding QGP medium.
The mass-dependent effect of the radial flow could give a larger $D_{s}/D^{0}$ ratio in heavy-ion collisions with respect to $p+p$ collisions.
The estimated $p_{T}$ dependence of the $D_{s}/D^{0}$ ratio using the Blast Wave parameters obtained from the $D^{0}$ spectra~\cite{HFTD0} shows a small difference at low $p_{T}$, and it increases to $\sim$10\% at 6 GeV/$c$ with respect to the ratio obtained from the PYTHIA calculation.    

The $p_{T}$-integrated $D_{s}$ yield is calculated by summing the data in the measured $p_{T}$ region and the estimated yield in the unmeasured region ($p_{T}$ $<$ 1.5 GeV/$c$). The latter is estimated as follows. 
For each centrality, the shape of the $D_{s}$ $p_{T}$ spectrum is obtained by multiplying the measured $D^{0}$ $p_{T}$ spectrum (parameterized by a Levy fit~\cite{Levy}) by the $D_{s}/D^{0}$ ratio from various model calculations. The normalization is fixed by a fit to the $D_{s}$ data points. 
The average of different fit functions is used to calculate the central value of the $D_{s}$ yield in the unmeasured $p_{T}$ region, and the maximum deviation of the yield estimated with the different shapes of the $D_{s}$ spectra is included in the systematic uncertainty.
The fractions of the extrapolated yield are 68\% for 0--10\% centrality, 42\% for 10--40\% centrality, and 65\% for 40--80\% centrality.
The $p_{T}$-integrated yields, $dN/dy$, are estimated to be 0.317 $\pm$ 0.038 (stat.) $\pm$ 0.11 (sys.) for 0--10\% centrality, 0.162 $\pm$ 0.017 (stat.) $\pm$ 0.042 (sys.) for 10--40\% centrality and 0.0202 $\pm$ 0.0016 (stat.) $\pm$ 0.0046 (sys.) for 40--80\% centrality.
The $p_{T}$-integrated $D_{s}/D^{0}$ yield ratio is 0.42 $\pm$ 0.04 (stat.) $\pm$ 0.11 (sys.) in 10--40\% centrality. The value estimated from THERMUS (a statistical hadronization model)~\cite{THERMUS}, employing a grand canonical ensemble with $T_{\rm ch}$ = 160 MeV, $\mu_{B}$ = 21.9 MeV and strangeness fugacity $\gamma_{s}$ = 1.0, is $\sim$0.36, consistent with the measured value within uncertainties.

\begin{figure}[htbp]
\center{
\includegraphics[width=0.9\columnwidth]{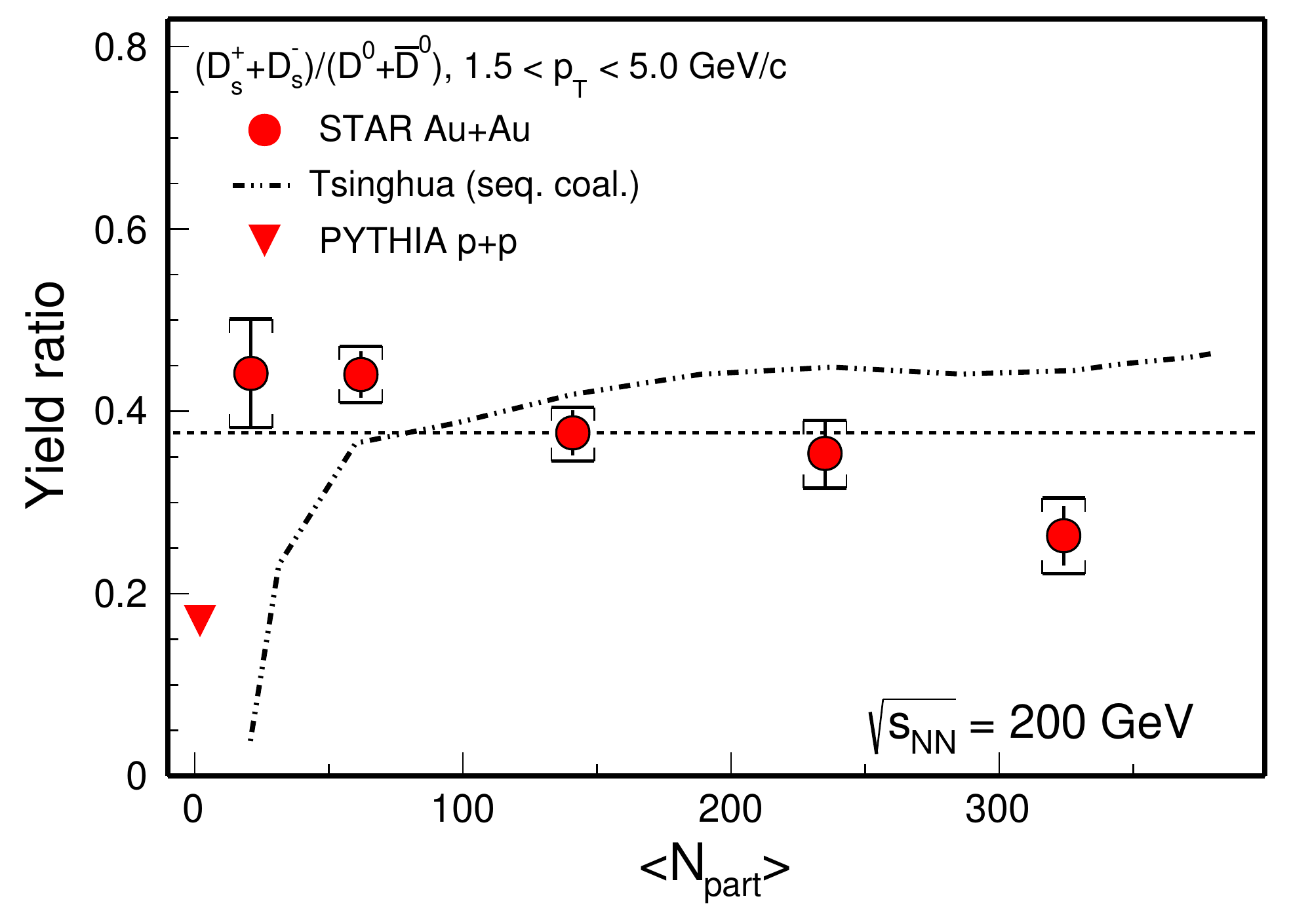}
}
\caption{ The integrated $D_{s}/D^{0}$ yield ratio (red solid circles) within 1.5 $<$ $p_{T}$ $<$ 5\,GeV/$c$ as a function of collision centrality (expressed as $\left \langle N_{\rm part}\right \rangle $) compared to Tsinghua model calculations (dash-dot-dot line) in Au+Au collisions at $\sqrt{s_{_{\rm NN}}}$ = 200\,GeV. The dashed line represents a fit of the $D_{s}/D^{0}$ data to a constant value. Vertical bars and brackets on data points represent statistical and systematic uncertainties, respectively.}
\label{fig:dNdyRatio}
\end{figure}

The $D_{s}/D^{0}$ ratios integrated over 1.5 $<$ $p_{T}$ $<$ 5\,GeV/$c$, as a function of the average number of participating nucleons ($N_{\rm part}$), are shown in Fig. 4.
A clear enhancement ($\sim$1.5-2.3 times) is found for the $p_{T}$-integrated $D_{s}/D^{0}$ ratio in Au+Au collisions compared to the value from PYTHIA in $p$+$p$ collisions at $\sqrt{s}$ = 200\,GeV.
The significances of the enhancement are 2.2, 5.1, 7.3, 8.6 and 4.5 standard deviations for 0--10\%, 10--20\%, 20--40\%, 40--60\%, and 60--80\% collision centralities, respectively. 
The calculation from Tsinghua (seq. coal.)~\cite{Tsinghua} underestimates the data from the most peripheral collisions, as also seen in the bottom panel of Fig. 3, and overestimates the data from central collisions. 

In summary, in this letter we presented the first measurement of $D_{s}^{\pm}$ production and $D_{s}/D^{0}$ yield ratio as a function of $p_{T}$, for different collision centralities at midrapidity ($|y|$\,$<$\,1) in Au+Au collisions at $\sqrt{s_{_{\rm NN}}}$ = 200\,GeV.
A clear enhancement of the $D_{s}/D^{0}$ yield ratio is found compared to PYTHIA simulations of $p$+$p$ collisions at the same collision energy.
For the $D_{s}/D^{0}$ ratios integrated over 1.5 $<$ $p_{T}$ $<$ 5\,GeV/$c$, in the 10--60\% centrality range, the significance of this observation is more than 5 standard deviations. 
The $p_{T}$-integrated $D_{s}/D^{0}$ ratio is compatible with the prediction from a statistical hadronization model.
The enhancement and its $p_{T}$ dependence can be qualitatively described by model calculations incorporating thermal abundance of strange quarks in the QGP and coalescence hadronization of charm quarks.
These results suggest that recombination of charm quarks with strange quarks in the QGP plays an important role in $D_{s}^{\pm}$ meson production in heavy-ion collisions.

We thank the RHIC Operations Group and RCF at BNL, the NERSC Center at LBNL, and the Open Science Grid consortium for providing resources and support.  This work was supported in part by the Office of Nuclear Physics within the U.S. DOE Office of Science, the U.S. National Science Foundation, the Ministry of Education and Science of the Russian Federation, National Natural Science Foundation of China, Chinese Academy of Science, the Ministry of Science and Technology of China and the Chinese Ministry of Education, the Higher Education Sprout Project by Ministry of Education at NCKU, the National Research Foundation of Korea, Czech Science Foundation and Ministry of Education, Youth and Sports of the Czech Republic, Hungarian National Research, Development and Innovation Office, New National Excellency Programme of the Hungarian Ministry of Human Capacities, Department of Atomic Energy and Department of Science and Technology of the Government of India, the National Science Centre of Poland, the Ministry  of Science, Education and Sports of the Republic of Croatia, RosAtom of Russia and German Bundesministerium fur Bildung, Wissenschaft, Forschung and Technologie (BMBF), Helmholtz Association, Ministry of Education, Culture, Sports, Science, and Technology (MEXT) and Japan Society for the Promotion of Science (JSPS).

\end{document}